\documentclass[prl,twocolumn,floatfix]{revtex4}
\usepackage{color}
\usepackage{epsfig}
\usepackage{graphicx}
\usepackage{amsmath}
\usepackage{array}
\newcommand {\be}  {
\begin{equation}
}
\newcommand {\ee}  {
\end{equation}
}
\newcommand {\bea} {
\begin{eqnarray}
}
\newcommand {\eea} {
\end{eqnarray}
}

\newcommand {\ma} {
\mathcal A
}

\newcommand {\mb} {
\mathcal B
}

\newcommand {\mc} {
 \mathcal C
}

\newcommand{\boldnabla}{\text{\boldmath$\nabla$}}

\newcommand{\m}{\mathcal}

\begin{document}

\title{\bf   Kinetic cross coupling between non-conserved and conserved fields in phase field models}
\author{Efim A. Brener$^1$ and G. Boussinot$^{1,2}$}
\affiliation{$^1$Peter Gr{\"u}nberg Institut, Forschungszentrum J{\"u}lich, D-52425 J{\"u}lich, Germany 
\\
$^2$Computational Materials Design Department, 
Max-Planck Institut f\"ur Eisenforschung, D-40074 D\"usseldorf, Germany}
\date{\today}
\begin{abstract}
We present a phase field model for
 isothermal transformations of  two component  alloys that includes Onsager kinetic cross coupling  between  the non-conserved phase field $\phi$ and the conserved concentration field $C$. We also provide  the reduction of the phase field model to the corresponding macroscopic description of the free boundary problem.  The reduction is given  in a general form. Additionally we  use an explicit example of a phase field model and check that the reduced macroscopic description, in the range of its applicability, is in excellent agreement with direct phase field simulations. The relevance of the newly introduced terms to solute trapping is also discussed.

\end{abstract}
\maketitle
{\it  Introduction.}
 Interface kinetics often plays a very important role in phase transformations. It is responsible for the deviation from the local thermodynamic equilibrium at the interface between different phases. In the case of alloys, it may affect the microstructure and the average concentration of the growing phase. It is responsible for solute trapping and some oscillatory instabilities of the transformation front leading eventually to the formation of banded structures. 

For example, in the case of isothermal transformations of binary alloys the bulk of each phase is described by the diffusion equation. The interface kinetics is more complicated because phase transformation effects occur in this region in addition to the diffusional exchange. In the macroscopic phenomenological approach to this problem (see, for example, \cite{caroli} and references therein) it is assumed that the physical interface  width is much smaller than any relevant macroscopic length scale and one formulates linear  Onsager relations at the interface which describe the interface kinetics. These relations connect two independent fluxes $J_A$ and $J_B$ (through the interface) of atoms  $A$ and $B$ to two independent driving forces $\delta\mu_A$ and $\delta\mu_B$ which are the differences in the chemical potentials of $A$ and $B$ atoms at the interface (see below).  The corresponding symmetric  positive-definite Onsager matrix   fully describes the interface kinetic properties in the framework of linear nonequilibrium thermodynamics. Of course, as in any phenomenological description, the three independent elements of this matrix depend on the  specific physical mechanisms  in the interface region. This means that any specific thermodynamically consistent model of the interface (in general nonlinear) being linearized near the equilibrium must be reducible to this phenomenological description and the elements of the Onsager matrix should be calculated in terms of the model parameters. 

In recent years the phase field approach to phase transformations has attracted the attention of much research (see, for example,  \cite{PE} and references therein). It was originally introduced as a mathematical tool to solve the free boundary problem without directly tracking the interface position. In the case of isothermal transformations of binary alloys, this approach introduces, in addition to the conserved concentration field $C$,  a nonconserved phase field $\phi$. This field changes smoothly on the scale of the interface width from some value, say $\phi=0$, that corresponds to one phase  to some other value, $\phi=1$, which corresponds to the other phase.  
The  phase field equations of motion  have a "diagonal" form in the classical variational formulation (see below),
i.e. the time derivative of $\phi$ ($C$) depends only on the functional derivative of the free energy with respect to $\phi$ ($C$). 
 This "diagonal" formulation therefore contains 
 only two independent coefficients describing the interface kinetic properties while the general macroscopic phenomenology allows three independent parameters.  An intuitively clear way to resolve this problem is to introduce kinetic cross coupling (non-diagonal terms) directly into the phase field equations. However, to  our best knowledge, this idea of a more general description of the interface kinetics in phase field models was never discussed in the literature (but see a very recent paper \cite{SZP} for a different approach). Moreover, 
as stated in \cite{Sekerka98}, according to Curie's principle \cite{GM62},  there can be no kinetic coupling between the scalar non-conserved phase field   $\phi$ and vectorial diffusional fluxes of the conserved quantities  energy and/or concentration. We think that this is an erroneous  statement (see also remarks in
\cite{BT}). The presence of the interface and the existence of the vector $\boldnabla\phi$,
that is orthogonal to the interface and operates only in the interface region,  
 allows  to formulate phase field equations  that include kinetic cross coupling  and are in agreement with linear nonequilibrium thermodynamics and Curie's principle.  This issue is also  relevant to the  anti-trapping current introduced in some non-variational versions of the phase field model \cite{K,KP} for different purposes. The anti-trapping current introduces  a new kinetic coefficient and uses $\boldnabla\phi$ as a vector normal to the interface.  To use this idea for the description of the cross effect of the interface kinetics in  phase field models, one should carefully consider  the necessary Onsager symmetry.  This goal can be achieved only  in the variational formulation of the phase field model. 

The main purpose of this paper is to provide a thermodynamically consistent  description of the phase field model which contains cross kinetic coupling between the phase field $\phi$ and the concentration field $C$. 
The second goal is to provide the reduction of this phase field model to the macroscopic phenomenological description as described above. This reduction is important to understand which macroscopic problem can be solved by the phase field model. 
The phase field model
contains explicitly the finite interface width $W$ as a parameter that is not included in the macroscopic description. We know only one example of such a reduction for classical ("diagonal") phase field models which keeps  $W$ finite (the sharp interface limit sets $W=0$). The  thin interface limit was originally introduced by Karma and Rappel \cite{KR} for the temperature field instead of the concentration field. 
This approach has then been promoted by many authors including the concentration field in the discussion (see, for example, a very detailed paper by Elder {\it et. al} \cite{E}).  However, Brener and Temkin \cite{BT} have recently pointed out  that the macroscopic description derived by the thin interface limit has a clear deficit in some range of parameters of the original phase field model. Namely, it can create strong unphysical instabilities due to a violation of the positive-definiteness of the obtained Onsager matrix, while the original phase field model is fully consistent and stable. 
The approach promoted in the present paper for the reduction to the macroscopic description is free from this deficit. 


{\it  Macroscopic description of isothermal alloy transformations.}
We discuss  the phase transformation of  a two component alloy at a given temperature $T$ with an interface separating two phases. The dimensionless concentration of B atoms is $C_1$ in   growing phase 1 and $C_2$ in  mother phase 2. In the bulk of each phase these concentrations are described by diffusion equations with diffusion coefficients $D_1$ and $D_2$. 
In order to formulate the boundary conditions at the interface we use the phenomenological Onsager approach. Onsager relations connect the  fluxes $J_A$ and $J_B$ (through the interface) of atoms  $A$ and $B$ to two driving forces $\delta\mu_A$ and $\delta\mu_B$ which are the usual differences in the chemical potentials of $A$ and $B$ atoms at the interface (see, for example, \cite{caroli} and references therein), 
\begin{eqnarray}
\label{mu3}
\delta\mu_A/T&=& \m  AJ_A+\m BJ_B \ ,\\
\label{mu4}
\delta\mu_B/T&=& \m BJ_A+\m CJ_B \ .
\end{eqnarray} 
The Onsager matrix should be positive-definite: $\m A$ and $\m C$ must be positive and $\m B^2<\m A\m C$.
According to  the conservation of  $B$ atoms  at the interface we also have \cite{T81, caroli}
\begin{eqnarray}
\label{cont1}
-D_1({\bf n}\cdot\boldnabla C_1)&=&VC_1-J_B \ ,\\
\label{cont2}
-D_2({\bf n}\cdot\boldnabla C_2)&=& VC_2-J_B\ , \\
\label{V}
                V&=&J_A+J_B \ ,
\end{eqnarray} 
 where ${\bf n}$ is the unit vector normal to the interface and $V$ is the normal velocity of the interface. 
 In this description the matrix of Onsager coefficients describes a positive entropy production (per unit area), 
$T\dot s=J_A\delta\mu_A+J_B\delta\mu_B,$ 
in the interface region. 
For the following it is useful to use $V=J_A+J_B$ and $J_B$ as independent fluxes and 
$\delta\mu_A$ and $\delta\mu=\delta\mu_B-\delta\mu_A$ as corresponding driving forces. This choice preserves the invariance of the entropy production, 
\begin{equation}
\label{entr1}
T\dot s=J_A\delta\mu_A+J_B\delta\mu_B=V\delta\mu_A+J_B\delta\mu \ .
\end{equation}
In this representation the linear relations between
driving forces and fluxes read:
\begin{eqnarray}
\label{mu5}
\delta\mu_A/T&=& \m{\bar A}V+ \m{\bar B}J_B \ ,\\
\label{mu6}
\delta\mu/T&=& \m {\bar B}V+ \m{ \bar C}J_B \ ,
\end{eqnarray} 
with $
\m A=\m{\bar A} \ , \m B=\m{\bar B}+\m{\bar A} \ $, and $ \m C= \m{\bar C} +\m{\bar A}+2\m{\bar B}\ .$
We note that if $f(C,T)$ is the free energy density of the phase, then often $\mu= \mu_B-\mu_A=\partial f/\partial  C$ is  called  the diffusion chemical potential and $\mu_A= f(C)-\mu C$  the grand potential.

{ \it  Phase field approach.}
We normalize the total free energy $F$  by $T$ and write the dimensionless free energy $G$ in the standard form for phase field models, 
\begin{equation}
G=F/T=\int dV \Big\{H \big[W^2(\boldnabla\phi)^2/2+f_{DW}(\phi) \big]+g(C,\phi)\Big\} .
\end{equation}
Here $f_{DW}=\phi^2(1-\phi^2)$ is the normalized double-well potential which has equal minima at $\phi=0$ and $\phi=1$; $W$ is the characteristic scale of the interface width; $g(C,\phi)$ is the dimensionless density of the chemical free energy;  $H$ represents the relative amplitude of the double-well potential normalized by $T$ and H is usually a large parameter. We assume that bulk phase $1$ corresponds to $\phi=1$ and bulk phase $2$ to $\phi=0$. 
Then, the dimensionless density of the bulk free energy is $g(C,1)=f_1(C)/T$ and $g(C,0)=f_2(C)/T$. The detailed form of $g(C,\phi)$ is model dependent.

We write the system of phase field equations in the following variational form: 

\begin{eqnarray}
\label{eq1}
-\delta G/\delta\phi&=& \tau\dot\phi+ M_{\phi}W({\bf J}\cdot\boldnabla\phi) \ ,\\
\label{eq2}
-\boldnabla(\delta G/\delta C)&=& M_CW\dot\phi \boldnabla\phi + {\bf J}/ D(\phi)\ ,\\
\label{cont}
\dot C+ (\boldnabla\cdot {\bf J})&=&0 \ .
\end{eqnarray} 
In the bulk of each phase only the ${\bf J}$ terms survive leading to the usual diffusional flux, ${\bf J}_1=-D_1\boldnabla C_1$ and
${\bf J}_2=-D_2\boldnabla C_2$,
with the bulk diffusion coefficients,  $D_1=[D(\phi=1)/T)]\partial \mu_1/\partial C$ and $D_2=[D(\phi=0)/T]\partial \mu_2/\partial C$.
In the interface region all terms are important leading to more complicated kinetics. 

The expression for the total entropy production reads 
\bea\label{entr2}
\dot S &=& \int dV [-\dot\phi\delta G/\delta\phi -{\bf J}\cdot\boldnabla(\delta G/\delta C )]   \\ 
&=&\int dV[\tau  \big(\dot\phi\big) ^2+{\bf J}^2/D(\phi)+ (M_{\phi}+M_C)W\dot\phi({\bf J}\cdot\boldnabla\phi)] \;.  \nonumber   
\eea
Onsager symmetry requires  
\begin{equation}
\label{sym}
M_{\phi}=M_C=M.
\end{equation}
The conditions of positive-definiteness of the entropy production read, $\tau>0$, $D(\phi_0)>0$ and 
\begin{equation}
\label{pos}
\tau/M^2> \text{max}[W^2D(\phi_0)(\boldnabla\phi_0)^2] \ ,
\end{equation}
where $\phi_0$ is the phase field distribution at thermodynamic equilibrium.

In classical phase field models $M=0$. The terms with $M$ represent kinetic cross coupling and introduce a new kinetic coefficient. 
 A term analogous to our  term with $M_C$ has been introduced in \cite{K,KP} for a different purpose, using non-variational versions of phase field equations. 
 To our best knowledge, the term with $M_{\phi}$ has never been included before in phase field theory. Moreover, this term must be included in a thermodynamically consistent  theory due to Onsager symmetry, Eq. (\ref{sym}), as soon as the $M_C$ term is included. 

 {\it Reduction of the phase field  description to the macroscopic description.} 
We integrate the phase  field equations over the interface region in order to derive effective boundary conditions in the form of Eqs. (\ref{mu5}) and (\ref{mu6}). This will allow us to express the macroscopic elements of the Onsager matrix in terms of the phase field parameters and to have an additional check of the symmetry condition.
 
 We assume that the interface is locally flat because we are mainly interested in kinetic effects rather than in the Gibbs-Thomson curvature correction and  denote the direction normal to the interface by $x$. In the interface region we make a quasi-stationnary approximation, $\dot\phi\approx -V\phi '(x)$ and $\dot C\approx -VC'(x)$, due to the strong gradients of $\phi$ and $C$ in this region even at thermodynamic equilibrium. 
 Integrating the continuity equation (\ref{cont}) in the interface region and choosing the integration constant  equal to $-J_B$   
 we find, 
 \begin{equation}
 \label{flux}
 J(x) \approx  -J_B+VC(x).  
\end{equation}
Eq. (\ref{flux}) then reproduces macroscopic continuity  equations, (\ref{cont1}) and (\ref{cont2}), if the observation point $x$ is chosen in phase $1$ or $2$ near the interface. We integrate Eq. (\ref{eq2}), 
\bea
\delta\mu/T &= & V\left [M_CW\int_W dx [\phi_0'(x)]^2-\int_Wdx \; C_0(x)/D(\phi_0)\right] \label{mu} \nonumber \\
 &+&J_B\int_Wdx/D(\phi_0) \;, 
\eea
and also, multiplying Eq. (\ref{eq1}) by $\phi'(x)$ and integrating over the same region, we find 
\bea\label{mua}
\delta\mu_A/T= V \Big[ \tau\int_W dx [\phi_0'(x)]^2+\int_Wdx \;C_0^2(x)/D(\phi_0) \nonumber \\
-(M_{\phi}+M_C)W\int_W dx [\phi_0'(x)]^2C_0(x) \Big] \nonumber \\
+J_B\left [M_{\phi}W\int_W dx [\phi_0'(x)]^2-\int_W \;dx \;C_0(x)/D(\phi_0)\right].
\eea
Here $\int_W$ denotes the integral over the interface region whose width is of order $W$, but such that $\phi$ ranges from $\phi \approx 1$ to $\phi \approx 0$.
We have replaced $\phi(x)$ and $C(x)$ by  their equilibrium distributions, $\phi_0(x)$ and $C_0(x)$, due to linearization. 
We have also used the following steps in order to integrate the left-hand-side of  Eq. (\ref{eq1}). First, the contribution proportional to $H$ vanishes \cite{explanation}. Secondly, we write 
$$\int_Wdx \;\phi'(x)\partial g/\partial\phi=\int_Wdg-\int_Wdx \; C'(x)\partial g/\partial C \;,$$
 integrate the last term by parts  
$$\int_Wdx \; \phi'(x)\partial g/\partial\phi=\delta\mu_A/T +\int_Wdx \;C(x)(\partial g/\partial C)' \;,$$
and use again Eq. (\ref{eq2}) for $(\partial g/\partial C)'$. 

As  expected Eq. (\ref{mua}) has the form of Eq. (\ref{mu5}) and  Eq. (\ref{mu}) has the form of Eq. (\ref{mu6}). The Onsager symmetry of Eqs. (\ref{mu5}) and (\ref{mu6}) requires 
$M_{\phi}=M_C=M$
that confirms Eq. (\ref{sym}). 
 We can also easily check that the interfacial part of the entropy production, Eq. (\ref{entr2}),  reduces 
to the form of Eq. (\ref{entr1}) with $\delta\mu$ given by Eq. (\ref{mu}) and $\delta\mu_A$ given by Eq. (\ref{mua}). It is of course positive if the condition (\ref{pos}) is fulfilled.
The phase field model presented here contains three independent inverse velocity scales describing the interface kinetics, $\tau/W$, $\int_Wdx/D(\phi) $ and $M$, while classical phase field models include only two. 

{\it Explicit example and numerical checks}. 
Our aim now is  to compare quantitatively simulation results within a specific phase-field model to the solution of the corresponding macroscopic description using the reduction presented above.
 We focus on the one-dimensional steady-state growth of phase 1 at the expense of phase 2, a case where the growth velocity $V$ is kinetically controlled. Due to the global conservation law, the concentration $C_1$ in phase 1 is constant, $C_1=C_{\infty}$,  where $C_\infty$ is the concentration far ahead of the interface in phase 2.  
 Within the macroscopic description, in the limit of small velocity, $V$ and $C_2$ read \cite{BT}
\bea\label{v_ons}
V &=& \frac{(f_1''(C_1^{eq})/T)
(C_1^{eq}-C_ \infty)\Delta C} {\bar  \ma + \bar  \mb (C_1^{eq}+C_2^{eq}) + \bar  \mc C_1^{eq} C_2^{eq}} \\
C_2 &=& C_2^{eq}+(C_ \infty  -C_1^{eq}) + (\bar  \mb + \bar  \mc C_1^{eq}) V \label{c2_ons}
\eea
where $f''_1(C)$ is the second derivative of $f_1(C)$ with respect to $C$ and $\Delta C = C_2^{eq}-C_1^{eq}$ with $C_1^{eq}$($C_2^{eq}$) the two-phase equilibrium concentration of phase 1(2).

We use a simple phase-field model for which the chemical free energy densities $f_1(C)$ and $f_2(C)$ of phases 1 and 2  parabolically depend on the concentration,  
\bea
 g(C,\phi) = \frac{1}{2} \Big(C - C_2^{eq} - p(\phi) \Delta C  \Big)^2
\eea
with $p(\phi) = \phi^3(10-15\phi+6\phi^2)$ (see for example \cite{FP}).
For an equilibrium interface centered at $x=0$, we have: 
$\phi_0(x) = 1/2 - \tanh [ x/(\sqrt{2} W) ]/2 $
with $\phi_0=1$ in phase 1 and $\phi_0=0$ in phase 2;
 $C_0(x) = (C_1^{eq}+C_2^{eq})/2 + u(x) \Delta C /2 $
with $u(x)=-u(-x)=1-2p[\phi_0(x)]$.

 
For simplicity, and in order to make further analytical progress, we assume a constant diffusion coefficient $D(\phi) = D$. 
This assumption is  physically more relevant to solid-solid transformations than to solidification problems where $D_1 \ll D_2$.  
 We perform the integrations in Eqs. (\ref{mu}, \ref{mua}) in a  symmetric range $[-\delta,\delta]$ around $x=0$ yielding:
\bea
\bar  \ma = && \frac{\alpha  \tau}{W}  -  \beta W \Delta C^2/(4D) \nonumber \\
&& + [({C_1^{eq} })^2+ {(C_2^{eq}})^2 ]   \delta/D   - M \alpha (C_1^{eq}+C_2^{eq})   \;,  \\
\bar  \mb =&& M\alpha - \left(C_1^{eq}+C_2^{eq}\right) \delta/D   \;, \\
\bar  \mc =&& 2\delta/D \;,
\eea 
where $\delta\sim W$ but such that  $\phi_0(-\delta)\approx1$ and $\phi_0(\delta)\approx 0$. Then
\bea
\alpha = W \int_{-\delta}^\delta dx \; [\phi_0'(x)]^2 \approx  W\int_{-\infty}^\infty dx \; [\phi_0'(x)]^2 
\approx 0.23570  \;, \nonumber \\
\beta =  \int_{-\delta}^ \delta  \frac{dx}{W} \; [1-u^2(x)]  \approx   \int_{-\infty}^\infty  \frac{dx}{W} \; [ 1-u^2(x)] 
 \approx  1.40748 \;, \nonumber
\eea
due to the fast convergence of the integrals. 
Using the latter kinetic coefficients in Eqs. (\ref{v_ons}) and (\ref{c2_ons}), we find $V$ and $C_2$ 
($f_1''(C_1^{eq})/T = 1$):
\bea \label{v_ph_field}
V &=&\frac{(C_1^{eq}-C_ \infty)\Delta C}{ \alpha  \tau/W -  \beta W  \Delta C^2/(4D)}  \;, \\
C_2 &=& C_2^{eq}+C_ \infty-C_1^{eq} + \left( M \alpha - \delta \Delta C /D \right) V . \label{c2_ph_field}
\eea
While the velocity is essentially independent of the integration range $\delta$ as discussed in \cite{BT}, the concentration $C_2$ depends on $\delta$. 
We note that in our reduced description  the interface concentrations and chemical potentials are actually defined at the spatial points $x=\pm\delta$ and vary  slightly  
with $\delta$ due to weak gradients if the system slightly deviates from equilibrium. For a more detailed discussion of this issue and its relation to the extrapolation procedure in the thin interface limit \cite{KR}, see \cite{BT}.


\begin{figure}[htbp]
\includegraphics[angle=0,width=230pt]{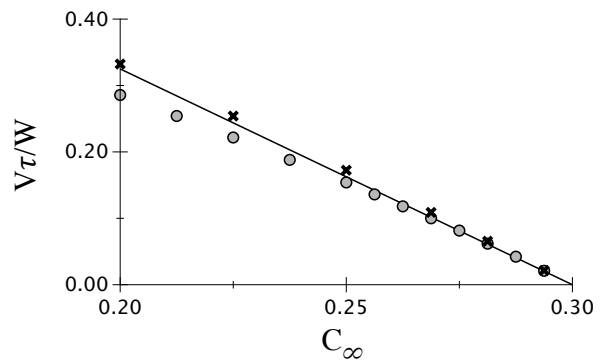}
\caption{\label{velocity} Dimensionless velocity $V\tau/W$ vs concentration of the system $C_\infty$ for different values of $M$ (crosses: $MW/\tau=2$; circles: $MW/\tau=0$) compared with the analytical prediction (line) of Eq. (\ref{v_ph_field}). The case $MW/\tau=-2$ is indistinguishable from $MW/\tau=2$.}
\end{figure}

\begin{figure}[htbp]
\includegraphics[angle=0,width=230pt]{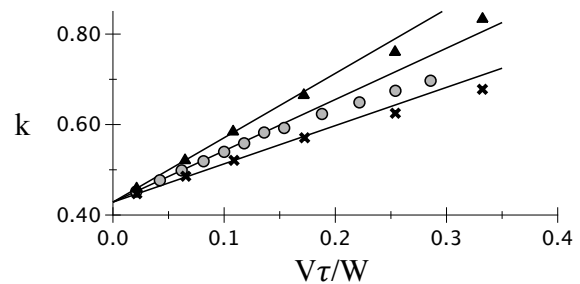}
\caption{\label{partition_coeff} Partition coefficient $k$ vs dimensionless velocity for different values of $M$ (crosses: $MW/\tau=2$; circles: $MW/\tau=0$; triangles: $MW/\tau =-2$) compared with the corresponding analytical prediction (lines) of Eq. (\ref{c2_ph_field}) with $\delta=2\sqrt{2} W$.}
\end{figure}

In Fig. \ref{velocity}, we present a comparison of the dimensionless velocity $V\tau/W$ as a function of $C_\infty$ given by the analytical formula,  Eq. (\ref{v_ph_field}),  and obtained from phase-field simulations. The two equilibrium concentrations are $C_1^{eq} = 0.3$ and $C_2^{eq} = 0.7$, the diffusion coefficient (constant throughout the whole system) is $D\tau/W^2=0.5$ and $H=50$ (we checked that the results are essentially independent of $H$ for such large values).
We find a good quantitative agreement in the linear regime, i.e. for small velocities. The simulations reproduce the independence of $M$ for the velocity in the linear regime (see Eq. (\ref{v_ph_field})). Nonlinearities  of the phase field model naturally lead to deviations  at higher velocities. 
We mention that here the denominator in Eq. (\ref{v_ph_field}) is positive. 
For smaller values of $D$, the denominator may be negative and steady-state solutions exist even for $C_\infty > C_1^{eq}$ (see, for example,  \cite{BT, L, KBS} and references therein).


In Fig. \ref{partition_coeff}, we present   the partition coefficient, $k=C_1/C_2$ (we recall that in steady-state $C_1=C_\infty$), as a function of the dimensionless velocity for different  values of $M$, with $C_2$ measured at $x=\delta = 2\sqrt{2}W$. The classical phase field model ($M=0$) shows already the solute trapping effect (increase of the partition coefficient with velocity) while positive values of $M$ show anti-trapping tendency and negative values of $M$ promote further solute trapping. 
This was the reason for authors of \cite{K,KP} to include a term with positive $M_C$ in Eq.(\ref{eq2}) calling it anti-trapping current. 
However, we understand now that a thermodynamically consistent description requires simultaneously to include the term with $M_{\phi}=M_C$ in  phase field Eq. (\ref{eq1}). 

We have also checked numerically the stability condition, Eq. (\ref{pos}) (for our explicit example it reads $8\tau/(DM^2)>1$), by investigating the relaxation to the equilibrium configuration. If the condition is violated by $1.5\%$ the system "blows up" instead of relaxing to the equilibrium. 

{\it{Summary.}} We have formulated a phase field model given by Eqs. (\ref{eq1}), (\ref{eq2}), and (\ref{cont}). It includes Onsager kinetic cross coupling  between  the non-conserved phase field $\phi$ and the conserved concentration field $C$. We have performed the reduction of this model to the corresponding macroscopic description given by Eqs. (\ref{mu}) and (\ref{mua}).


{\it Acknowledgment}. 
We are grateful to D.E. Temkin for useful discussions. We acknowledge the support of the Deutsche Forschungs- gemeinschaft under Project  SFB 917.


\begin{thebibliography}{99}
\bibitem{explanation} This is the case because we consider a locally flat interface. For a curved interface, the integral which involves the Laplacian of $\phi$ leads to the classical Gibbs-Thomson curvature correction which is proportional to the interface free energy. Moreover we do not discuss surface diffusion and interface stretching (see for example \cite{K, KP}) which are generically smaller than interface kinetic effects in the macroscopic limit.
\end{thebibliography}
\end{document}